\documentclass[12pt]{article}
\usepackage{epsfig,amsfonts,amssymb}
\usepackage{hyperref}
\pdfoutput=1

\usepackage{cite}
\usepackage{cite}

\usepackage{comment}

\def\ZZZ{{\hbox{ Z\kern-1.6mm Z}}}
\def\RRR{{\hbox{ R\kern-2.4mm R}}}
\def\CCC{{\hbox{ C\kern-2.0mm C}}}
\def\zzz{{\hbox{z\kern-1mm z}}}

\newcommand{\qeq}{{\hbox{=\kern-2.3mm ? \kern.5mm }}}
\renewcommand{\qeq}{=}

\newcommand{\eps}{\epsilon}

\newcommand{\ve}{\varepsilon}

\newcommand{\MM}{{\cal M}}

\newcommand{\OO}{{\cal O}}

\newcommand{\be}{\begin{equation}}
\newcommand{\ee}{\end{equation}}
\newcommand{\ben}{\begin{eqnarray}\displaystyle}
\newcommand{\een}{\end{eqnarray}}

\newcommand{\refb}[1]{(\ref{#1})}

\newcommand{\sectiono}[1]{\section{#1}\setcounter{equation}{0}}

\def\one{{\hbox{ 1\kern-.8mm l}}}
\def\zero{{\hbox{ 0\kern-1.5mm 0}}}

\newcommand{\bea}[1]{\begin{eqnarray}\label{#1} }
\newcommand{\eea}{\end{eqnarray}}

\newcommand{\eqref}{\refb}

\usepackage{bm}
\usepackage[table]{xcolor}

\def\figone{

\def\JPicScale{0.8}
\ifx\JPicScale\undefined\def\JPicScale{1}\fi
\unitlength \JPicScale mm
\begin{picture}(200,70)(0,0)
\linethickness{0.3mm}
\put(0,50){\line(1,0){20}}

\put(18,70){\makebox(0,0)[cc]{$\vdots\, m$}}

\put(53,72){\makebox(0,0)[cc]{$\vdots\, p$}}

\put(100,72){\makebox(0,0)[cc]{$\vdots$}}

\put(125,70){\makebox(0,0)[cc]{$\vdots$}}

\put(180,70){\makebox(0,0)[cc]{$\vdots$}}

\linethickness{0.3mm}
\put(40,50){\line(1,0){20}}
\linethickness{0.3mm}
\put(80,50){\line(1,0){20}}
\linethickness{0.3mm}
\put(120,50){\line(1,0){20}}
\linethickness{0.3mm}
\put(160,50){\line(1,0){20}}
\linethickness{0.3mm}
\put(180,50){\line(1,0){20}}
\linethickness{0.3mm}
\qbezier(20,50)(25.19,55.25)(30,55.25)
\qbezier(30,55.25)(34.81,55.25)(40,50)
\linethickness{0.3mm}
\qbezier(20,50)(25.19,44.75)(30,44.75)
\qbezier(30,44.75)(34.81,44.75)(40,50)
\linethickness{0.3mm}
\linethickness{0.3mm}
\linethickness{0.3mm}
\qbezier(20,50)(14.77,55.2)(13.56,58.81)
\qbezier(13.56,58.81)(12.36,62.42)(15,65)
\qbezier(15,65)(17.61,67.67)(18.81,64.06)
\qbezier(18.81,64.06)(20.02,60.45)(20,50)
\linethickness{0.3mm}
\qbezier(40,50)(45.2,60.45)(48.81,64.06)
\qbezier(48.81,64.06)(52.42,67.67)(55,65)
\qbezier(55,65)(57.67,62.42)(54.06,58.81)
\qbezier(54.06,58.81)(50.45,55.2)(40,50)
\linethickness{0.3mm}
\qbezier(100,50)(105.19,44.75)(110,44.75)
\qbezier(110,44.75)(114.81,44.75)(120,50)
\linethickness{0.3mm}
\qbezier(100,50)(105.19,55.25)(110,55.25)
\qbezier(110,55.25)(114.81,55.25)(120,50)
\linethickness{0.3mm}
\put(31,51){\makebox(0,0)[cc]{$\vdots\, n$}}
\linethickness{0.3mm}
\put(110,51){\makebox(0,0)[cc]{$\vdots$}}
\linethickness{0.3mm}
\qbezier(100,50)(94.75,55.2)(94.75,58.81)
\qbezier(94.75,58.81)(94.75,62.42)(100,65)
\qbezier(100,65)(105.25,67.67)(105.25,64.06)
\qbezier(105.25,64.06)(105.25,60.45)(100,50)
\linethickness{0.3mm}
\qbezier(120,50)(119.98,55.22)(121.19,57.62)
\qbezier(121.19,57.62)(122.39,60.03)(125,60)
\qbezier(125,60)(127.64,60.03)(126.44,57.62)
\qbezier(126.44,57.62)(125.23,55.22)(120,50)
\linethickness{0.3mm}
\qbezier(180,50)(174.75,55.22)(174.75,57.62)
\qbezier(174.75,57.62)(174.75,60.03)(180,60)
\qbezier(180,60)(185.25,60.03)(185.25,57.62)
\qbezier(185.25,57.62)(185.25,55.22)(180,50)
\linethickness{0.3mm}
\put(20,50){\makebox(0,0)[cc]{{\Large $\bullet$}}}

\put(40,50){\makebox(0,0)[cc]{{\Large $\circ$}}}

\put(100,50){\makebox(0,0)[cc]{{\Large $\circ$}}}

\put(120,50){\makebox(0,0)[cc]{{\Large $\bullet$}}}

\put(180,50){\makebox(0,0)[cc]{$\square$}}

\put(0,45){\makebox(0,0)[cc]{1}}

\put(60,45){\makebox(0,0)[cc]{2}}

\put(80,45){\makebox(0,0)[cc]{1}}

\put(140,45){\makebox(0,0)[cc]{2}}

\put(160,45){\makebox(0,0)[cc]{1}}

\put(200,45){\makebox(0,0)[cc]{2}}

\put(30,40){\makebox(0,0)[cc]{(a)}}

\put(110,40){\makebox(0,0)[cc]{(b)}}

\put(180,40){\makebox(0,0)[cc]{(c)}}

\linethickness{0.3mm}
\qbezier(20,50)(12.16,55.2)(9.75,59.41)
\qbezier(9.75,59.41)(7.34,63.62)(10,67.5)
\qbezier(10,67.5)(12.59,71.42)(15,72.62)
\qbezier(15,72.62)(17.41,73.83)(20,72.5)
\qbezier(20,72.5)(22.62,71.26)(22.62,65.84)
\qbezier(22.62,65.84)(22.62,60.43)(20,50)
\linethickness{0.3mm}
\qbezier(40,50)(39.97,65.66)(42.38,72.88)
\qbezier(42.38,72.88)(44.78,80.09)(50,80)
\qbezier(50,80)(55.2,80.02)(58.81,78.22)
\qbezier(58.81,78.22)(62.42,76.41)(65,72.5)
\qbezier(65,72.5)(67.62,68.6)(67.62,65.59)
\qbezier(67.62,65.59)(67.62,62.59)(65,60)
\qbezier(65,60)(62.45,57.41)(56.44,55)
\qbezier(56.44,55)(50.42,52.59)(40,50)
\linethickness{0.3mm}
\qbezier(100,50)(89.55,55.2)(85.34,59.41)
\qbezier(85.34,59.41)(81.13,63.62)(82.5,67.5)
\qbezier(82.5,67.5)(83.78,71.41)(86.19,73.81)
\qbezier(86.19,73.81)(88.59,76.22)(92.5,77.5)
\qbezier(92.5,77.5)(96.38,78.82)(100.59,78.22)
\qbezier(100.59,78.22)(104.8,77.62)(110,75)
\qbezier(110,75)(115.28,72.45)(112.88,66.44)
\qbezier(112.88,66.44)(110.47,60.42)(100,50)
\linethickness{0.3mm}
\qbezier(120,50)(117.38,60.42)(117.38,66.44)
\qbezier(117.38,66.44)(117.38,72.45)(120,75)
\qbezier(120,75)(122.59,77.62)(125,77.62)
\qbezier(125,77.62)(127.41,77.62)(130,75)
\qbezier(130,75)(132.6,72.41)(134.41,70)
\qbezier(134.41,70)(136.21,67.59)(137.5,65)
\qbezier(137.5,65)(138.87,62.42)(134.66,58.81)
\qbezier(134.66,58.81)(130.45,55.2)(120,50)
\linethickness{0.3mm}
\qbezier(180,50)(169.55,55.19)(165.94,60)
\qbezier(165.94,60)(162.33,64.81)(165,70)
\qbezier(165,70)(167.58,75.22)(171.19,77.62)
\qbezier(171.19,77.62)(174.8,80.03)(180,80)
\qbezier(180,80)(185.2,80.03)(188.81,77.62)
\qbezier(188.81,77.62)(192.42,75.22)(195,70)
\qbezier(195,70)(197.67,64.81)(194.06,60)
\qbezier(194.06,60)(190.45,55.19)(180,50)

\end{picture}

}

\def\figtwo{

\def\JPicScale{0.8}
\ifx\JPicScale\undefined\def\JPicScale{1}\fi
\unitlength \JPicScale mm
\begin{picture}(200,70)(0,0)
\linethickness{0.3mm}
\put(0,50){\line(1,0){20}}

\put(15,70){\makebox(0,0)[cc]{$\vdots$}}

\put(55,72){\makebox(0,0)[cc]{$\vdots$}}

\put(100,72){\makebox(0,0)[cc]{$\vdots$}}

\put(125,70){\makebox(0,0)[cc]{$\vdots$}}

\linethickness{0.3mm}
\put(20,30){\line(0,1){20}}
\linethickness{0.3mm}
\put(100,30){\line(0,1){20}}
\linethickness{0.3mm}
\put(80,50){\line(1,0){20}}
\linethickness{0.3mm}
\linethickness{0.3mm}
\linethickness{0.3mm}
\qbezier(20,50)(25.19,55.25)(30,55.25)
\qbezier(30,55.25)(34.81,55.25)(40,50)
\linethickness{0.3mm}
\qbezier(20,50)(25.19,44.75)(30,44.75)
\qbezier(30,44.75)(34.81,44.75)(40,50)
\linethickness{0.3mm}
\linethickness{0.3mm}
\linethickness{0.3mm}
\qbezier(20,50)(14.77,55.2)(13.56,58.81)
\qbezier(13.56,58.81)(12.36,62.42)(15,65)
\qbezier(15,65)(17.61,67.67)(18.81,64.06)
\qbezier(18.81,64.06)(20.02,60.45)(20,50)
\linethickness{0.3mm}
\qbezier(40,50)(45.2,60.45)(48.81,64.06)
\qbezier(48.81,64.06)(52.42,67.67)(55,65)
\qbezier(55,65)(57.67,62.42)(54.06,58.81)
\qbezier(54.06,58.81)(50.45,55.2)(40,50)
\linethickness{0.3mm}
\qbezier(100,50)(105.19,44.75)(110,44.75)
\qbezier(110,44.75)(114.81,44.75)(120,50)
\linethickness{0.3mm}
\qbezier(100,50)(105.19,55.25)(110,55.25)
\qbezier(110,55.25)(114.81,55.25)(120,50)
\linethickness{0.3mm}
\put(30,51){\makebox(0,0)[cc]{$\vdots$}}
\linethickness{0.3mm}
\put(110,51){\makebox(0,0)[cc]{$\vdots$}}
\linethickness{0.3mm}
\qbezier(100,50)(94.75,55.2)(94.75,58.81)
\qbezier(94.75,58.81)(94.75,62.42)(100,65)
\qbezier(100,65)(105.25,67.67)(105.25,64.06)
\qbezier(105.25,64.06)(105.25,60.45)(100,50)
\linethickness{0.3mm}
\qbezier(120,50)(119.98,55.22)(121.19,57.62)
\qbezier(121.19,57.62)(122.39,60.03)(125,60)
\qbezier(125,60)(127.64,60.03)(126.44,57.62)
\qbezier(126.44,57.62)(125.23,55.22)(120,50)

\linethickness{0.3mm}
\put(20,50){\makebox(0,0)[cc]{{\Large $\bullet$}}}

\put(40,50){\makebox(0,0)[cc]{{\Large $\circ$}}}

\put(100,50){\makebox(0,0)[cc]{{\Large $\circ$}}}

\put(120,50){\makebox(0,0)[cc]{{\Large $\bullet$}}}

\put(0,45){\makebox(0,0)[cc]{1}}

\put(23,30){\makebox(0,0)[cc]{2}}

\put(98,30){\makebox(0,0)[cc]{2}}

\put(80,45){\makebox(0,0)[cc]{1}}

\put(30,40){\makebox(0,0)[cc]{(d)}}

\put(110,40){\makebox(0,0)[cc]{(e)}}

\linethickness{0.3mm}
\qbezier(20,50)(12.16,55.2)(9.75,59.41)
\qbezier(9.75,59.41)(7.34,63.62)(10,67.5)
\qbezier(10,67.5)(12.59,71.42)(15,72.62)
\qbezier(15,72.62)(17.41,73.83)(20,72.5)
\qbezier(20,72.5)(22.62,71.26)(22.62,65.84)
\qbezier(22.62,65.84)(22.62,60.43)(20,50)
\linethickness{0.3mm}
\qbezier(40,50)(39.97,65.66)(42.38,72.88)
\qbezier(42.38,72.88)(44.78,80.09)(50,80)
\qbezier(50,80)(55.2,80.02)(58.81,78.22)
\qbezier(58.81,78.22)(62.42,76.41)(65,72.5)
\qbezier(65,72.5)(67.62,68.6)(67.62,65.59)
\qbezier(67.62,65.59)(67.62,62.59)(65,60)
\qbezier(65,60)(62.45,57.41)(56.44,55)
\qbezier(56.44,55)(50.42,52.59)(40,50)
\linethickness{0.3mm}
\qbezier(100,50)(89.55,55.2)(85.34,59.41)
\qbezier(85.34,59.41)(81.13,63.62)(82.5,67.5)
\qbezier(82.5,67.5)(83.78,71.41)(86.19,73.81)
\qbezier(86.19,73.81)(88.59,76.22)(92.5,77.5)
\qbezier(92.5,77.5)(96.38,78.82)(100.59,78.22)
\qbezier(100.59,78.22)(104.8,77.62)(110,75)
\qbezier(110,75)(115.28,72.45)(112.88,66.44)
\qbezier(112.88,66.44)(110.47,60.42)(100,50)
\linethickness{0.3mm}
\qbezier(120,50)(117.38,60.42)(117.38,66.44)
\qbezier(117.38,66.44)(117.38,72.45)(120,75)
\qbezier(120,75)(122.59,77.62)(125,77.62)
\qbezier(125,77.62)(127.41,77.62)(130,75)
\qbezier(130,75)(132.6,72.41)(134.41,70)
\qbezier(134.41,70)(136.21,67.59)(137.5,65)
\qbezier(137.5,65)(138.87,62.42)(134.66,58.81)
\qbezier(134.66,58.81)(130.45,55.2)(120,50)

\end{picture}

}

\def\figfive{

\def\JPicScale{0.8}
\ifx\JPicScale\undefined\def\JPicScale{1}\fi
\unitlength \JPicScale mm
\begin{picture}(110,75)(0,0)
\linethickness{0.3mm}
\put(70,0){\line(0,1){30}}
\linethickness{1mm}
\multiput(30,70)(0.12,-0.12){333}{\line(1,0){0.12}}
\linethickness{1mm}
\multiput(70,30)(0.12,0.12){333}{\line(1,0){0.12}}
\linethickness{0.3mm}
\put(60,40){\line(0,1){30}}
\linethickness{0.3mm}
\put(50,50){\line(0,1){20}}
\linethickness{0.3mm}
\put(40,60){\line(0,1){10}}
\linethickness{0.3mm}
\put(80,40){\line(0,1){30}}
\linethickness{0.3mm}
\put(90,50){\line(0,1){20}}
\linethickness{0.3mm}
\put(100,60){\line(0,1){10}}
\put(45,65){\makebox(0,0)[cc]{$\cdots$}}

\put(95,65){\makebox(0,0)[cc]{$\cdots$}}

\put(80,75){\makebox(0,0)[cc]{$e_1$}}

\put(90,75){\makebox(0,0)[cc]{$e_2$}}

\put(100,75){\makebox(0,0)[cc]{$e_r$}}

\put(60,75){\makebox(0,0)[cc]{$e_1'$}}

\put(50,75){\makebox(0,0)[cc]{$e_2'$}}

\put(40,75){\makebox(0,0)[cc]{$e_l'$}}

\put(110,75){\makebox(0,0)[cc]{R}}

\put(110,65){\makebox(0,0)[cc]{$e$}}

\put(30,75){\makebox(0,0)[cc]{L}}

\put(30,65){\makebox(0,0)[cc]{$e'$}}

\put(67,28){\makebox(0,0)[cc]{$P$}}

\put(52,33){\makebox(0,0)[cc]{$e'+\sum_i e'_i$}}

\put(87,33){\makebox(0,0)[cc]{$e+\sum_i e_i$}}

\end{picture}

}

\def\figthree{

\def\JPicScale{0.8}
\ifx\JPicScale\undefined\def\JPicScale{1}\fi
\unitlength \JPicScale mm
\begin{picture}(200,70)(0,0)
\linethickness{0.3mm}
\put(0,50){\line(1,0){20}}

\linethickness{0.1mm}

\put(30,40){\line(0,1){20}}

\put(110,40){\line(0,1){20}}

\linethickness{0.3mm}

\put(15,70){\makebox(0,0)[cc]{$\vdots$}}

\put(55,72){\makebox(0,0)[cc]{$\vdots$}}

\put(100,72){\makebox(0,0)[cc]{$\vdots$}}

\put(125,70){\makebox(0,0)[cc]{$\vdots$}}

\linethickness{0.3mm}
\linethickness{0.3mm}
\linethickness{0.3mm}
\put(80,50){\line(1,0){20}}
\linethickness{0.3mm}
\linethickness{0.3mm}
\linethickness{0.3mm}
\qbezier(20,50)(25.19,55.25)(30,55.25)
\qbezier(30,55.25)(34.81,55.25)(40,50)
\linethickness{0.3mm}
\qbezier(20,50)(25.19,44.75)(30,44.75)
\qbezier(30,44.75)(34.81,44.75)(40,50)
\linethickness{0.3mm}
\linethickness{0.3mm}
\linethickness{0.3mm}
\qbezier(20,50)(14.77,55.2)(13.56,58.81)
\qbezier(13.56,58.81)(12.36,62.42)(15,65)
\qbezier(15,65)(17.61,67.67)(18.81,64.06)
\qbezier(18.81,64.06)(20.02,60.45)(20,50)
\linethickness{0.3mm}
\qbezier(40,50)(45.2,60.45)(48.81,64.06)
\qbezier(48.81,64.06)(52.42,67.67)(55,65)
\qbezier(55,65)(57.67,62.42)(54.06,58.81)
\qbezier(54.06,58.81)(50.45,55.2)(40,50)
\linethickness{0.3mm}
\qbezier(100,50)(105.19,44.75)(110,44.75)
\qbezier(110,44.75)(114.81,44.75)(120,50)
\linethickness{0.3mm}
\qbezier(100,50)(105.19,55.25)(110,55.25)
\qbezier(110,55.25)(114.81,55.25)(120,50)
\linethickness{0.3mm}
\put(40,50){\line(1,0){20}}
\put(33,51){\makebox(0,0)[cc]{$\vdots$}}
\linethickness{0.3mm}
\put(120,50){\line(1,0){20}}
\put(113,51){\makebox(0,0)[cc]{$\vdots$}}
\linethickness{0.3mm}
\qbezier(100,50)(94.75,55.2)(94.75,58.81)
\qbezier(94.75,58.81)(94.75,62.42)(100,65)
\qbezier(100,65)(105.25,67.67)(105.25,64.06)
\qbezier(105.25,64.06)(105.25,60.45)(100,50)
\linethickness{0.3mm}
\qbezier(120,50)(119.98,55.22)(121.19,57.62)
\qbezier(121.19,57.62)(122.39,60.03)(125,60)
\qbezier(125,60)(127.64,60.03)(126.44,57.62)
\qbezier(126.44,57.62)(125.23,55.22)(120,50)

\linethickness{0.3mm}
\put(20,50){\makebox(0,0)[cc]{{\Large $\bullet$}}}

\put(40,50){\makebox(0,0)[cc]{{\Large $\circ$}}}

\put(100,50){\makebox(0,0)[cc]{{\Large $\circ$}}}

\put(120,50){\makebox(0,0)[cc]{{\Large $\bullet$}}}

\put(0,45){\makebox(0,0)[cc]{1}}

\put(60,45){\makebox(0,0)[cc]{2}}

\put(140,45){\makebox(0,0)[cc]{2}}

\put(80,45){\makebox(0,0)[cc]{1}}

\put(30,30){\makebox(0,0)[cc]{(a)}}

\put(110,30){\makebox(0,0)[cc]{(b)}}

\linethickness{0.3mm}
\qbezier(20,50)(12.16,55.2)(9.75,59.41)
\qbezier(9.75,59.41)(7.34,63.62)(10,67.5)
\qbezier(10,67.5)(12.59,71.42)(15,72.62)
\qbezier(15,72.62)(17.41,73.83)(20,72.5)
\qbezier(20,72.5)(22.62,71.26)(22.62,65.84)
\qbezier(22.62,65.84)(22.62,60.43)(20,50)
\linethickness{0.3mm}
\qbezier(40,50)(39.97,65.66)(42.38,72.88)
\qbezier(42.38,72.88)(44.78,80.09)(50,80)
\qbezier(50,80)(55.2,80.02)(58.81,78.22)
\qbezier(58.81,78.22)(62.42,76.41)(65,72.5)
\qbezier(65,72.5)(67.62,68.6)(67.62,65.59)
\qbezier(67.62,65.59)(67.62,62.59)(65,60)
\qbezier(65,60)(62.45,57.41)(56.44,55)
\qbezier(56.44,55)(50.42,52.59)(40,50)
\linethickness{0.3mm}
\qbezier(100,50)(89.55,55.2)(85.34,59.41)
\qbezier(85.34,59.41)(81.13,63.62)(82.5,67.5)
\qbezier(82.5,67.5)(83.78,71.41)(86.19,73.81)
\qbezier(86.19,73.81)(88.59,76.22)(92.5,77.5)
\qbezier(92.5,77.5)(96.38,78.82)(100.59,78.22)
\qbezier(100.59,78.22)(104.8,77.62)(110,75)
\qbezier(110,75)(115.28,72.45)(112.88,66.44)
\qbezier(112.88,66.44)(110.47,60.42)(100,50)
\linethickness{0.3mm}
\qbezier(120,50)(117.38,60.42)(117.38,66.44)
\qbezier(117.38,66.44)(117.38,72.45)(120,75)
\qbezier(120,75)(122.59,77.62)(125,77.62)
\qbezier(125,77.62)(127.41,77.62)(130,75)
\qbezier(130,75)(132.6,72.41)(134.41,70)
\qbezier(134.41,70)(136.21,67.59)(137.5,65)
\qbezier(137.5,65)(138.87,62.42)(134.66,58.81)
\qbezier(134.66,58.81)(130.45,55.2)(120,50)

\end{picture}

}

\begin{document}

\baselineskip 24pt

\begin{center}

{\Large \bf Infrared finite semi-inclusive cross section in two dimensional type 0B
string theory}

\end{center}

\vskip .6cm
\medskip

\vspace*{4.0ex}

\baselineskip=18pt

\centerline{\large \rm Ashoke Sen}

\vspace*{4.0ex}

\centerline{\large \it International Centre for Theoretical Sciences - TIFR 
}
\centerline{\large \it  Bengaluru - 560089, India}


\vspace*{1.0ex}
\centerline{\small E-mail:  ashoke.sen@icts.res.in}

\vspace*{5.0ex}

\centerline{\bf Abstract} \bigskip

D-instanton induced S-matrix in type 0B string theory in two dimensions suffers from infrared
divergences. This can be traced to the fact that these processes produce low energy
rolling tachyon states that cannot be regarded as linear combination of finite number of
closed string states. We compute
semi-inclusive cross sections in this theory where we allow in the final state
a fixed set of closed strings carrying given 
energies and any number of other closed string states
carrying the rest of the energy.
The result is infrared finite and agrees with the results in the
dual matrix model, described by non-relativistic fermions moving in an inverted
harmonic oscillator potential. In the matrix model
the role of `any number of other closed string states'
is played by a fermion hole pair on opposite sides of the potential barrier.

\vfill \eject

\tableofcontents

\sectiono{Introduction and summary} \label{s1}

Non-critical string theories in two dimensions provide toy models for critical string theory
where many of the computational tools in string theory can be tested\cite{dj,sw,gk,kr}.
Of these, type 0B
string theory provides a non-perturbatively consistent string
theory\cite{0307083,0307195,2201.05621}. Its dual matrix model description is a theory of free
fermions moving in an
inverted harmonic oscillator potential, with energy levels filled up to a fermi level that lies
below the maximum of the potential. The closed string states represent excitations
of the fermi sea involving low energy fermion hole pairs,
with the parity even excitations describing
NS sector states and parity odd excitations describing R-sector states.
On the other hand, single fermions of the matrix model
represent rolling tachyon configurations\cite{0203211,0203265} on unstable D-branes
of the theory\cite{0304224,0305159,0307083,0307195,0307221}.

 In this theory, perturbative amplitudes for external closed strings involve reflection
 of the fermion hole pair excitations near the fermi sea from the potential barrier. The
 transmission through the barrier are non-perturbative in the string coupling, and represent
 D-instanton effects. The effect of a single D-instanton is to transmit a single fermion or a
 single hole across the barrier. The corresponding final state cannot be interpreted as a 
 conventional closed string since it involves a pair of fermion and hole on {\it opposite
 sides of the potential barrier}. In the string theory computation of closed string S-matrix, this
 is reflected in the fact that single D-instanton (or anti-D-instanton) mediated processes are
 infrared divergent\cite{0309148}.
 In fact these divergences exponentiate and make the amplitude vanish.
 On the other hand, a D-instanton - anti-D-instanton induced process involves either the
 transmission of a fermion and a hole across the barrier, or a non-perturbative 
 contribution to the reflection amplitude of a fermion or a hole. In either case, the final
 state represents fermion hole pair excitations on the same side of the barrier, and therefore
 can be interpreted as a regular closed string state, On the string theory side this is
 reflected in the fact that the amplitudes induced by a D-instanton anti-D-instanton pair are
 infrared finite, Explicit computation of these amplitudes yield results in perfect agreement
 with the predictions of the matrix model\cite{2204.01747,2207.07138}.
 
 In this paper we analyze the single (anti-) D-instanton induced 
 amplitudes in more detail. In particular we show that even though the infrared divergences
 in the exponent make the S-matrix of a fixed set of external closed string states
 vanish, we have infrared
 finite semi-inclusive cross section where we sum over
 final states containing a fixed number of closed strings with given energies and 
 arbitrary number of other closed string states carrying the rest of the energy. This can
 then be compared to the matrix model result for a similar semi-inclusive cross section.
 However in the matrix model computation, we can replace the 
 `arbitrary number of other closed string states'  in the final state
by  an additional
 fermion hole pair on opposite sides of the potential barrier, since single (anti-) D-instanton
 induced processes produce such fermion hole pair.
 The agreement between the two computations
 suggests that single fermion or hole excitations
 in the matrix model, representing rolling tachyon configuration on the 
 unstable D0-brane of type 0B string theory, can be regarded as a collection of infinite
 number of closed strings, as expected from the rules of bosonization.
 
 For convenience of the reader, we shall now describe our main result. Instead of
 working with NSNS and RR sector states, we work with right and left sector states, given
 respectively by the sum and difference of the NSNS and RR sector states. In the matrix model
 language these are represented by fermion hole pair excitations on the right and the
 left of the potential barrier. We start with a single incoming right sector closed string carrying
 energy $\omega_1$ and compute the semi-inclusive cross section for producing a final
 state containing $r$  right sector closed
string states of energies in the range $(e_1, e_1+\Delta e_1),\cdots, 
(e_r, e_r+\Delta e_r)$,
$l$ left sector closed
string states of energies in the range $(e'_1, e'_1+\Delta e'_1),\cdots, 
(e'_l, e_l+\Delta e'_l)$
and any number of other closed string states. If we denote a final state satisfying these
requirements by $\langle n|$ and denote by $\MM_1(\omega_1,n)$ the transition
amplitude for this process, then for infinitesimal $\Delta e_i$, $\Delta e_i'$ we have
\ben\label{ebegin}
&& {\sum_n}' \MM_1(\omega_1,n) \MM_1(\omega_2,n)^* \nonumber \\
&=& \left\{\prod_{i=1}^r {\Delta e_i\over e_i} \right\}\, \left\{\prod_{i=1}^l {\Delta e'_i\over e'_i} \right\}\, 
\delta\left(\omega_1 -\omega_2 \right) {1\over \pi} 
\sinh\left(2\pi \left(\omega_1
- \sum_{i=1}^r e_i -\sum_{i=1}^l e'_i\right)\right) \nonumber \\ && \hskip 1in \times
\cosh\left(2\pi \left(\omega_1+ \sum_{i=1}^l e'_i - \sum_{i=1}^r e_i\right)\right)\, ,
\een
where the $'$ on the sum on the left hand side
is a reminder that we sum over only a restricted set of final
states.
Some salient features of this formula are as follows:
\begin{enumerate}
\item \refb{ebegin} represents the contribution to the semi-inclusive cross section due to
a single D-instanton or a single anti-D-instanton. As long as $l\ge 1$, i.e.\ the final
state contains at least one left sector closed string, this is the dominant contribution
to the cross section. However for $l=0$ there is also a perturbative contribution, not
shown here, that
dominates the result.
\item Even though the final formula \refb{ebegin}
is free from infrared divergence, the intermediate
steps of the calculation in string theory suffer from infrared divergences. We regulate the
infrared divergences by putting a lower cut-off on the spatial momentum. The matrix
model side of the calculation is free from infrared divergence at all steps. The difference
can be traced to the fact that in the matrix model we use the free fermion - hole basis for 
the part of the final state representing `any number of other closed string states'. This
allows us to include in the final state fermion and hole states on opposite sides of the
potential barrier. In contrast, a finite number of
closed string states in string theory describes only fermion
hole pairs on the same side of the potential.
\item In string theory, free fermion and hole states are represented by rolling tachyon
solution on unstable D0-brane\cite{0304224,0305159,0307083,0307195,0307221}. 
The agreement between the string theory and
matrix model results for the semi-inclusive cross section
suggests that we should be able to represent
these rolling tachyon configurations as infinite collection of closed strings.
\item Single (anti-) D-instanton contribution to the total cross section,
where we sum over all final states, is given by setting
$l=r=0$ in \refb{ebegin}, and yields a finite answer. This was already computed in
\cite{2204.01747}.
\item Naively one might expect that if we integrate \refb{ebegin} over the final state
energies $e_i$ and $e_i'$ and divide the result by the symmetry factor $l!r!$,
we shall get part of the total cross section that has at least
$l$ left sector closed string states and $r$ right sector closed string states
carrying any energy. However this is infrared divergent from
the $e_i\simeq 0$ and / or $e_i'\simeq 0$ region and would contradict the finiteness of
the total cross section. The resolution of this puzzle is provided by the fact that if the
final state had $p$ additional left sector closed strings and $q$ additional right
sector closed strings, then the computation of the total
cross section should include a factor of $1 / \{ (p+l)!(q+r)!\}$. However \refb{ebegin}
only includes a factor of $1/\{p!q!\}$. Therefore simple integration of \refb{ebegin} and
division by $l!r!$ will
overestimate the actual result by a factor of ${p+l\choose l}{q+r\choose r}$. For $p\to\infty$
or $q\to\infty$, this is an infinite factor.
\end{enumerate}

 The rest of the paper is organized as follows. In section \ref{s2} we review recent
 results of \cite{2204.01747,2207.07138}
 on D-instanton corrections to two dimensional type 0B string theory amplitudes.
 In section \ref{s3} we study the unitarity of the D-instanton anti-D-instanton induced
 amplitude. This requires studying the single (anti-) D-instanton
 induced amplitudes and regulating the infrared divergences in these amplitudes.
 This analysis was done earlier in \cite{2204.01747} using
 dimensional regularization. We use a lower cut-off on the spatial momentum
 to regulate the infrared divergences and also formulate the problem using the language
 of string field theory that makes the validity of Cutkosky rules and unitarity manifest.
 This also allows us to generalize the analysis to compute semi-inclusive cross sections.
 This is carried
out explicitly in section \ref{s4}, yielding the result \refb{ebegin}.
In section \ref{s5} we compute the same semi-inclusive
 cross section in the matrix model and show that the result agrees with the string theory
 results, even though the sum over states that we use in the matrix model looks different
 from the sum over states that we perform in string theory. We end in section \ref{s6} by 
 speculating on possible application to quantum electrodynamics in four dimensions.
 
\sectiono{Review} \label{s2}

The world-sheet theory of non-critical type 0B string theory in two dimensions has
a scalar describing the time direction, its world-sheet superpartner Majorana fermion,
super-Liouville theory with central charge $27/2$ and the usual $b,c,\beta,\gamma$ ghost
system. Physical closed string states in this theory are two scalars $\phi_{NS}$ and
$\phi_{R}$  from the NSNS sector and  the RR sector respectively. We shall denote their
vertex operators by $V_{NS}$, $V_R$ and work in the
$\alpha'=2$ unit as in \cite{2204.01747}.

This theory is expected to be dual to a matrix model. The simplest description of this model
is provided by the theory of non-interacting, non-relativistic fermions, each moving under
an inverted harmonic oscillator potential, 
and the energy levels are filled up to a fermi level
that is a height $\mu$ below the maximum of the
potential\cite{0307083,0307195,2201.05621}.
$\mu$ is inversely proportional to the string
coupling $g_s$. The asymptotic closed string
states are incoming and outgoing fermion hole pairs on the right
of the potential and the left of the potential. These can be identified to the
fields $\chi_R$ and
$\chi_L$ given by
sum and difference of $\phi_{NS}$ and $\phi_R$\cite{0307083,0307195,2201.05621}. 
We shall call them right and left sector closed string states respectively and, following
\cite{2204.01747},
normalize them such that
their vertex operators are given by $W_R= V_{NS}+V_R$ and $W_L = V_{NS}-V_R$
respectively. 
The parity symmetry of the inverted harmonic oscillator potential
translates to the $(-1)^{F_L}$
symmetry of the type 0B string theory under which the RR sector states change sign.

Let us consider the 2-point amplitude where the incoming and outgoing states 
are right sector closed strings of energy $\omega_1$ and $\omega_2$ respectively.
Instanton - anti-instanton contribution to this amplitude is given by\cite{2204.01747}:
\ben \label{e2.1}
\MM_2(\omega_1,\omega_2) &=&  e^{-2 S_D}
\exp\left[ {\int_0^\infty}{dt\over 2t} \left\{ -2 + 2\, e^{2 \pi t  \left({1\over 2} - {1\over 2}
\left({\Delta x\over 2\pi}\right)^2\right) }\right\}
\right]  \nonumber \\ &&\hskip -.5in \left(e^{\pi P_1 + i\omega^E_1 x_1} - e^{-\pi P_1
+i\omega^E_1 x_2}\right)
\left(e^{\pi P_2-i\omega^E_2 x_2} - e^{-\pi P_2-i\omega^E_2 x_1}\right),
\quad \omega^E \equiv -i\omega\, ,
\een
where $x_1$, $x_2$ are the positions of the D-instanton along the Euclidean time
direction and $\Delta x =x_1-x_2$.
Here $P_i$ denotes the Liouville momentum carried by the $i$-th particle and the
on-shell condition is $\omega_i=P_i$. $S_D$ is the action of a single D-instanton,
so that the $e^{-2S_D}$ term can be regarded as the result of summing over arbitrary
number of disk partition function with either D-instanton or anti-D-instanton boundary
condition. The last factor of the first line represents the exponential of the annulus
partition function, with the first term inside the curly bracket representing the contribution
from the annulus with both boundaries on the instanton or both boundaries on the
anti-instanton and the second term inside the curly bracket representing the contribution
from the annulus with one boundary lying on the instanton and the other boundary
lying on the anti-instanton. The first factor
in the second line is the contribution of the disk one point function of $W_L$ associated
with the incoming state, with the two terms representing the contribution from the disks
with instanton and anti-instanton boundary conditions respectively. Similarly the 
second factor
in the second line is the contribution of the disk one point function of $W_L$ associated
with the outgoing state, with the two terms representing the contribution from the disks
with anti-instanton and instanton boundary conditions respectively.

For fixed $x_1,x_2$,
the integral over $t$ has no divergence in the $t\to 0$ limit since the term inside the
curly bracket vanishes in this limit. 
The divergence at large $t$ are associated with the open string channel and
can be resolved using open string field theory. This gives a finite
result\cite{2207.07138}:\footnote{Note
that the computation in \cite{2207.07138} was done in the $\alpha'=1$ unit. The result quoted in
\refb{e2.2} is the translation of that result to the $\alpha'=2$ unit.}
\be\label{e2.2}
\exp\left[ {\int_0^\infty}{dt\over 2t} \left\{ -2 + 2\, e^{2 \pi t  \left({1\over 2} - {1\over 2}
\left({\Delta x)\over 2\pi}\right)^2\right) }\right\}\right]
= {1\over 4\pi^2} \int dx_1 dx_2 \, {1\over (\Delta x)^2 - 4\pi^2}\, ,
\ee
with the understanding that the integration over the zero modes $x_1,x_2$ should be
done at the end, after including the contribution from the disk amplitudes given in the
last line of \refb{e2.1}. Furthermore,
unitarity demands that we use the principal value prescription for dealing with the
singularities at $\Delta x =\pm 2\pi$\cite{2012.00041}. 
Substituting \refb{e2.2} into \refb{e2.1} we get\cite{2204.01747}
\ben
\MM_2(\omega_1,\omega_2) &=& e^{-2 S_D}
{1\over 4\pi^2} \int dx_1 dx_2 \, {1\over (x_1-x_2)^2 - 4\pi^2} \nonumber \\ && (e^{\pi P_1 + i\omega^E_1 x_1} - e^{-\pi P_1
+i\omega^E_1 x_2})
(e^{\pi P_2-i\omega^E_2 x_2} - e^{-\pi P_2-i\omega^E_2 x_1})\, .
\een
We can perform the integration over $x_1,x_2$ by changing variables to the center
of mass coordinate $x_1+x_2$ and the relative coordinate $\Delta x=x_1-x_2$.
The integration over $\Delta x$ may be done by closing the contour
at infinity in the upper / lower half plane. While doing this we need to keep in mind
that the analytic continuation from Euclidean energy $\omega^E$ to the Lorentzian
energy $\omega=i\omega^E$ has to be done via the first or third quadrant of the
complex $\omega$ plane. Therefore positive $\omega_1,\omega_2$ requires us to
start with positive $\omega^E_1$, $\omega^E_2$. After doing the integration over
$\Delta x$, we can analytically continue the energies to Lorentzian values
$\omega_1,\omega_2$. During this analytic continuation we also rotate the contour
of integration over $x_1+x_2$ so as to keep the combination $\omega x $ fixed.
At the end we are left with the integration over the center of mass coordinate along
the real time axis, and this integral  produces a factor of 
$2\pi i\delta(\omega_1
-\omega_2)$. The $i$ arises from having to express the integration over the center
of mass location along the imaginary time axis in terms of integration along the real 
time axis. The final result is\cite{2204.01747}
\be \label{e2.4aa}
\MM_2(\omega_1,\omega_2) = - e^{-2 S_D} {1\over 2\pi} \, \delta(\omega_1-\omega_2) \, \cosh(2\pi\omega_1)
\sinh(2\pi\omega_1)\, .
\ee

At this order we also have contribution from the
two instanton processes and two anti-instanton processes, but  these vanish due to
infrared divergence in the closed string channel ($t\to 0$ limit in the integral
in the exponent). For this reason we shall not consider these contributions. 
Similarly the contribution to the closed string S-matrix
due to single D-instanton or single anti-D-instanton also vanish due to infrared
divergence in the exponent of the normalization constant. Therefore the instanton -
anti-instanton contribution is apparently
the leading {\em instanton contribution} to the closed string
S-matrix, and we write,
\be \label{e2.5aa}
S(\omega_1,\omega_2)=\omega_1 \delta(\omega_1-\omega_2)+\MM_2(\omega_1,
\omega_2) \, .
\ee
Note that we have ignored the perturbative contribution to the S-matrix since
they will not play any role in our analysis. The $\omega_1$ in the definition of the identity
matrix on the right hand side indicates that the sum over states is performed with the
integration measure $d\omega/\omega$. We shall see in section \ref{s4} that this is the
correct choice for the normalization of the states that we have chosen.

We now compute a particular contribution to $S^\dagger S$:
\ben
\int {d\omega\over \omega} S(\omega_2,\omega)^* S(\omega_1,\omega)
&=& \int {d\omega \over \omega} \, 
\delta (\omega_1-\omega)  \, \left[\omega_1-  e^{-2 S_D} {1\over 2\pi} \, 
\cosh(2\pi\omega_1)
\sinh(2\pi\omega_1)\right]\nonumber\\
&&\delta (\omega_2-\omega)  \, \left[\omega_2-  e^{-2 S_D} {1\over 2\pi } \, 
 \cosh(2\pi\omega)
\sinh(2\pi\omega)\right]\nonumber\\
&& \hskip -1.5in =\, \delta (\omega_1-\omega_2)  \, \left[\omega_1- e^{-2 S_D} {1\over \pi} \, 
\cosh(2\pi\omega_1)
\sinh(2\pi\omega_1)+\OO\left(e^{-4 S_D}\right)\right]\, .
\een
Since this is not $\omega_1 \delta(\omega_1-\omega_2)$, the S-matrix is
apparently non-unitary.
The perturbative contribution is unitary by itself and cannot help cancel this term. The
interference term between $\MM_2$ and the perturbative S-matrix has additional powers
of string coupling and cannot contribute to this order. 
Therefore there must be additional contribution that has not been accounted for. The 
natural candidate is the contribution from single instanton or single anti-instanton. 
Even though we have argued that they vanish due to infrared divergences, let us
tentatively denote by $\MM_1(\omega,n)$ the single instanton (and single anti-instanton)
contribution to the S-matrix for transition from a 
closed string state of energy $\omega$ to an arbitrary state $n$. Then unitarity demands that
\be \label{e2.7}
\sum_n \MM_1(\omega_2,n)^* \MM_1(\omega_1,n) =\delta (\omega_1-\omega_2)  \, 
e^{-2 S_D} {1\over \pi} \, 
\cosh(2\pi\omega_1)
\sinh(2\pi\omega_1)\, .
\ee
Since we have seen that the D-instanton or anti-D-instanton induced contribution to the closed
string scattering amplitude vanishes, this poses an apparent conflict with
unitarity\cite{9111035,0309148}. The resolution to the 
puzzle is simplest in the matrix model. There closed strings are represented by fermion-hole
pair created on the same side of the potential, but the (anti-) D-instanton induced processes
create a fermion hole pair on opposite sides of the potential\cite{0309148}. Therefore,
without including these in the final state we should not expect to get a unitary S-matrix.
In string theory these are
represented by low energy rolling tachyon configurations on unstable
D0-branes\cite{0304224,0305159,0307083,0307195}.
This suggests that
in the sum over $n$ on the left hand side of \refb{e2.7} we must include these states
besides the closed string states in order to restore unitarity.

This however is not the end of the story. With the unitary prescription for integrating over
$\Delta x$, which in this case corresponds to using principal value prescription for integrating
across the singularity at $\Delta x=\pm 2\pi$, the S-matrix in the closed string sector
was shown to be unitary\cite{2012.00041}.
This will be in apparent conflict with the vanishing of $\MM_1$
in the closed string sector. We shall see in  section \ref{s3} that \refb{e2.7} holds for the
closed string S-matrix if we include in the set $n$ the states with infinite number of low
energy closed strings, without needing to sum over rolling tachyon states.
This was already checked in \cite{2204.01747} using a dimensional regularization scheme to
regulate the infrared divergence of $\MM_1$. We shall use a lower cut-off on the Liouville
momentum to regulate the infrared divergences and formulate the analysis in the
language of string field theory which makes the proof of unitarity manifest by relating it to
Cutkosky rules. This will also make the necessity of the principal value prescription
for integration over $\Delta x$ clear.
The result of this analysis can be interpreted as the statement that the
rolling tachyon state can be regarded as a state made of infinite number of closed strings.
The right hand side of \refb{e2.7} now gives the single instanton contribution to the
total cross section for a single closed string of energy $\omega_1$ to scatter to
any set of closed strings. The
situation is very similar to what happens in quantum electrodynamics. There the probability
of producing a set of charged states and a finite number of photons during a scattering
process vanishes due to infrared divergences.
However the inclusive cross section where we sum over all final states
is non zero and is consistent with unitarity\cite{bloch,yennie}.

\sectiono{Feynman diagram representation and Cutkosky rules} \label{s3}

In order to understand how the Cutkosky rules lead to \refb{e2.7},  
we shall first
formulate the computation of $\MM_2$ given in
section \ref{s2} as a sum of Feynman diagrams of string field theory. Once
this is done, the cuts of these Feynman diagrams, that keep the D-instanton induced
vertex and the anti-D-instanton induced vertex on two sides of the cut, will give the
contributions to the left hand side of \refb{e2.7}. On the other hand, the sum over cuts where
the D-instanton and the anti-D-instanton induced vertices are on the same side of the
cut, will give $\MM_2+\MM_2^*$.  Since Cutkosky rules tell us that the sum over all the cuts of a diagram
vanish, and since a general proof of this in a class of non-local theories that include
(effective) string field theory was given in \cite{1604.01783}, we are led to 
\refb{e2.7}. We shall also explicitly sum over the cut diagrams to
verify \refb{e2.7}. This will set up the framework for computing semi-inclusive cross
section.

Explicit check of \refb{e2.7} was carried out
in \cite{2204.01747} where the authors use dimensional
regularization scheme. Here we shall regularize the infrared divergences in the closed
string channel by putting a sharp lower cut-off on the Liouville momentum. As explained
above, the language
of string field theory that we shall use will make unitarity manifest following the analysis
of \cite{2012.00041}.

We express $\MM_2(\omega_1,\omega_2)$ given in \refb{e2.1} as,
\ben \label{e3.9}
\MM_2(\omega_1,\omega_2) &=& e^{-2 S_D}
\exp\Bigg[ {\int_\eps^\infty}{dt\over 2t} \left\{ -2 + 2 e^{-\pi t h}\right\}  - {\int_\eps^\infty}{dt\over t} 
\, e^{-\pi t h} \nonumber \\ &&
+{\int_0^\infty}{dt\over t} 
\left\{ e^{2 \pi t  \left({1\over 2} - {1\over 2}
\left({\Delta x\over 2\pi}\right)^2\right) } - \Theta(\eps-t)\right\}
\Bigg]  \nonumber \\ && (e^{\pi P_1 + i\omega^E_1 x_1} - e^{-\pi P_1
+i\omega^E_1 x_2})
(e^{\pi P_2-i\omega^E_2 x_2} - e^{-\pi P_2-i\omega^E_2 x_1})\,,
\een
for some small positive
number $\eps$ and an arbitrary positive number $h$. $\Theta$ is the
Heaviside step function.
Note that the $t$ integral in the second line diverges for $(\Delta x)^2\le 4\pi^2$.
We need to define the integral for $(\Delta x)^2 > 4\pi^2$ and then analytically continue
the result to $(\Delta x)^2<4\pi^2$, averaging over the contributions where $\Delta x$ goes
around the singularity at $\pm 2\pi$ above and below the singularity in the complex plane.
As explained in \cite{2012.00041},
this corresponds to a particular choice of integration contour in the
path integral over open string fields, since $x_1$ and $x_2$ are modes of the open string.
String field theory {\it a priori} does not fix the choice of contour, but a 
different choice of integration contour will lead to non-unitary amplitudes. Indeed, even an
otherwise good quantum field theory can be made bad if in the path integral over fields we
decide to integrate along a wrong choice of contour in the complex field space.

It follows from the analysis of \cite{2207.07138} that
up to corrections of order $\eps$, the contribution from
the first integral in the exponent of \refb{e3.9}
is given by replacing $\left({\Delta x\over 2\pi}\right)^2 -1$ by $h$
in \refb{e2.2}:
\be\label{e3.10}
\exp\left[ {\int_\eps^\infty}{dt\over 2t} \left\{ -2 + 2 e^{-\pi t h}\right\}\right]
= {1\over 16\pi^4\, h} \int dx_1 dx_2 +\OO(\eps) \, ,
\ee
with the understanding that the integrations over $x_1,x_2$ are to be performed after
including the rest of the contribution.
The contribution from the second integral in the exponent of \refb{e3.9} is given by,
\be\label{e3.11}
\exp\left[  - {\int_\eps^\infty}{dt\over t}
\, e^{-\pi t h} \right]  = \exp\left[\gamma_E + \ln(\pi\eps h) +\OO(\eps) \right] 
= \pi\, \eps \, h \, e^{\gamma_E} \left( 1 +\OO(\eps)\right)\, ,
\ee
where $\gamma_E$ is the Euler constant.
In the last integral in the exponent of \refb{e3.9} we change variable to $s=1/(2t)$
and write this as
\ben \label{e3.12}
&& \exp\left[{\int_0^\infty}{dt\over t}  \left\{ e^{2 \pi t  \left({1\over 2} - {1\over 2}
\left({\Delta x\over 2\pi}\right)^2\right) } - \Theta(\eps-t)\right\}\right]
\nonumber \\
&=& \exp\left[{\int_0^\infty}{ds\over s}  \left\{ e^{\pi s^{-1}  \left({1\over 2} - {1\over 2}
\left({\Delta x\over 2\pi}\right)^2\right) } - \Theta\left(s - {1\over 2\eps}\right)\right\}\right]
 \\
&=& \exp\left[4{\int_0^\infty}{ds}  \int {d^2 k_E} \, e^{- 2\pi s k_E^2}
\left\{ e^{-i\omega^E\Delta x} (\cosh^2 (\pi P) + \sinh^2(\pi P)) 
-  \Theta\left(s - {1\over 2\eps}\right) \right\} \right]\, , \nonumber
\een
where $k_E=(\omega^E, P)$ with $-\infty<\omega^E<\infty$, $0\le P<\infty$. 
It is easy to see that after doing the integral
over $k_E$, we reproduce the expression in the second line of \refb{e3.12}. 
Physically
$\omega^E$ represents Euclidean energy and
$P$ represents Liouville momentum.
We can now
exchange the order of integration and do the $s$ integral to write,
\ben\label{e3.13}
&& \exp\left[{\int_0^\infty}{dt\over t}  \left\{ e^{2 \pi t  \left({1\over 2} - {1\over 2}
\left({\Delta x\over 2\pi}\right)^2\right) } - \Theta(\eps-t)\right\}\right] \nonumber \\
&=& \exp\left[{2\over \pi} \int {d^2 k_E\over  k_E^2} 
\left\{ e^{-i\omega^E\Delta x} (\cosh^2 (\pi P) + \sinh^2(\pi P)) 
-  e^{-\pi k_E^2 /\eps} \right\} \right]\,.
\een
Note that if in \refb{e3.13} we try to carry out the integral over $P$ first,
the integral diverges for large $P$
due to the presence of the $(\cosh^2 (\pi P) + \sinh^2(\pi P))$ term
in the integrand. Since the original expression that we started with was finite for
$(\Delta x)^2 > 4\pi^2$, this divergence can be attributed to the exchange of the
order of integration over $s$ and $k_E$. This problem can be avoided if we follow the
prescription that the integration over $\omega^E$ has to be done before the integration
over $P$. In that case we can easily check, via closing the contour at infinity in the
complex $\omega^E$ plane and picking up residues at $\omega^E=\pm iP$ for negative /
positive $\Delta x$, that the
result of $\omega^E$ integration produces a factor proportional to $e^{-|\Delta x| P}$.  The
$P$ integral now converges for $|\Delta x|>2\pi$ and produces the original result.
Therefore from now on it will be understood that the $\omega^E$ integration needs to
be done before integration over $P$.

We can now substitute \refb{e3.10}, \refb{e3.11} and \refb{e3.13} into \refb{e3.9} to write
\ben\label{e3.14pre}
\MM_2(\omega_1,\omega_2) &=& e^{-2 S_D} {\eps\over 16\pi^3} e^{\gamma_E}
\int dx_1 dx_2 \nonumber \\ && 
\exp\left[ {2\over \pi}\int {d^2 k_E\over  k_E^2}  
\left\{ e^{-i\omega^E (x_1-x_2)} \left(\cosh^2 (\pi P) + \sinh^2(\pi P)\right) 
- e^{-\pi k_E^2 /\eps} \right\} \right] \nonumber \\ &&
(e^{\pi P_1 + i\omega^E_1 x_1} - e^{-\pi P_1
+i\omega^E_1 x_2})
(e^{\pi P_2-i\omega^E_2 x_2} - e^{-\pi P_2-i\omega^E_2 x_1}) \, .
\een
The proof of equivalence of \refb{e3.9} and \refb{e3.14pre} holds for real $\omega_1^E,
\omega_2^E$, i.e. imaginary $\omega_1, \omega_2$. In this case the integration
contour over the momenta $k=(\omega, P)=(i\omega^E, P)$ are taken to be along
real $\omega^E$, i.e. imaginary $\omega$ axis; $P$ is aways kept real and positive. 
To compute $\MM_2(\omega_1,\omega_2)$
for real $\omega_1,\omega_2$, we need to deform the external energies from the
imaginary axis to the real axis via the first quadrant of the complex $\omega_i$ plane.
During this deformation the poles of the propagators may approach the integration contours
and
we need to deform the integration contours over the internal energies $\omega=i\omega^E$ 
to avoid the poles, keeping the end-points fixed at $\pm i\infty$
so as to make use of the $e^{-\pi k^2/\eps}=e^{-\pi k_E^2/\eps}$ 
factor to make the integral converge at large
momentum\cite{1604.01783}.  Using the relation $d\omega^E=-id\omega$, we can replace
$d^2 k_E/k_E^2$ by $-i d^2 k/(k^2-i\ve)$  where the $i\ve$ in the denominator
essentially encodes the contour deformation prescription described above\cite{1604.01783}.

Note that  \refb{e3.14pre}
is free from infrared divergences in the closed
string channel, i.e. free from
divergences from the $k_E\simeq 0$ region. Therefore if we put a
lower limit $\eta$ on the integration range of the Liouville momentum $P$, then
the result is finite in the $\eta\to 0$ limit.
But now we can split the integral into sum
of terms each of which could diverge in the $\eta\to 0$ limit, manipulate them appropriately
and then combine the results before taking the $\eta\to 0$ limit.
With the understanding that we have a lower cut-off $\eta$ on $P$,
we rewrite \refb{e3.14pre} as,
\ben\label{e3.14}
\MM_2(\omega_1,\omega_2) &=& e^{-2 S_D} {\eps\over 16\pi^3} e^{\gamma_E}
\int dx_1 dx_2 \nonumber \\ && 
\exp\left[{2\over \pi} \int {d^2 k_E\over   k_E^2} e^{-\pi k_E^2 /\eps} 
 e^{-i\omega^E (x_1-x_2)} \left\{\cosh^2 (\pi P) + \sinh^2(\pi P)\right\} \right] \nonumber \\ &&
\exp\left[ -{1\over \pi}\int {d^2 k_E\over   k_E^2} e^{-\pi k_E^2 /\eps} 
\right] \exp\left[ -{1\over \pi}\int {d^2 k_E\over   k_E^2} e^{-\pi k_E^2 /\eps} 
\right] \nonumber \\ &&
\exp\left[ {2\over \pi} \int {d^2 k_E\over  k_E^2} \left(1-e^{-\pi k_E^2 /\eps} \right)
 e^{-i\omega^E (x_1-x_2)} \left\{\cosh^2 (\pi P) + \sinh^2(\pi P)\right\}
 \right] \nonumber \\ &&
(e^{\pi P_1 + i\omega^E_1 x_1} - e^{-\pi P_1
+i\omega^E_1 x_2})
(e^{\pi P_2-i\omega^E_2 x_2} - e^{-\pi P_2-i\omega^E_2 x_1}) \, .
\een

We can give this an interpretation in terms of Feynman diagrams by introducing a set of
D-instanton induced  vertices in the effective closed string
field theory\cite{2012.00041}. In writing down the
expressions for these vertices, we shall use both the Euclidean momenta
$k_E=(\omega^E,P)$
and the Lorentzian momenta $k=(\omega,P)$ with the understanding that $\omega=i\omega^E$.
\begin{enumerate}
\item
Single D-instanton induced $n$-point vertex {\Large $\bullet$} with external closed strings of
momenta $k_1=(\omega_1,P_1),
\cdots , k_n=(\omega_n, P_n)$:
\be\label{e3.15}
2\pi \, i\, \delta \left(\sum_{i=1}^n \sigma_i \omega_i\right) 
\left( e^{-2 S_D} {\eps\over 16\pi^3} e^{\gamma_E}\right)^{1/2} 
\prod_{i=1}^n e^{-\pi k_i^2 / (2\eps)} 
{\sigma_i \cosh\pi P_i\choose \sinh \pi P_i}\, ,
\ee
where $\cosh(\pi P)$ refers to RR-sector states, $\sinh(\pi P)$ refers to NSNS sector
states and $\sigma_i$ takes value $+1$ if the $i$-th state is incoming and $-1$ if the 
$i$-th state is outgoing.
\item
Single anti-D-instanton induced $n$-point 
vertex {\Large $\circ$} with external closed strings of
momenta $k_1=(\omega_1,P_1),
\cdots , k_n=(\omega_n, P_n)$:
\be\label{e3.16}
2\pi \, i\, \delta \left(\sum_{i=1}^n \sigma_i \omega_i\right) 
\left(e^{-2 S_D} {\eps\over 16\pi^3} e^{\gamma_E}\right)^{1/2}
\prod_{i=1}^n e^{-\pi k_i^2 / (2\eps)} 
{-\sigma_i \cosh\pi P_i\choose \sinh \pi P_i}\, .
\ee
\item
D-instanton - anti-D-instanton induced composite $n$-point vertex $\square$
with external closed strings of
momenta $k_1=(\omega_1,P_1),
\cdots , k_n=(\omega_n, P_n)$:
\ben\label{e3.17}
&& \int dx_1 dx_2 \prod_{i=1}^n e^{-\pi k_i^2 / (2\eps)} 
{\sigma_i \cosh\pi P_i (e^{i x_1 \sigma_i \omega_i^E} - e^{i x_2 \sigma_i
\omega_i^E}) \choose \sinh \pi P_i
(e^{i x_1 \sigma_i\omega_i^E} + e^{i x_2 \sigma_i \omega_i^E}) } \,
e^{-2 S_D} {\eps\over 16\pi^3} e^{\gamma_E} \nonumber \\ &&
\left[\exp\left\{ {2\over \pi} \int {d^2 k_E\over k_E^2} \left(1-e^{-\pi k_E^2 /\eps} \right)
\left\{ e^{-i\omega^E (x_1-x_2)} (\cosh^2 (\pi P) + \sinh^2(\pi P)) 
\right\} \right\}-1\right]\nonumber \\
&=& 2\pi\, i\, \delta\left(\sum_{i=1}^n \sigma_i \omega_i\right)\,
e^{-2 S_D} {\eps\over 16\pi^3} e^{\gamma_E}\nonumber \\ &&
\int d(\Delta x) \prod_{i=1}^n e^{-\pi k_i^2 / (2\eps)} 
{\sigma_i \cosh\pi P_i (e^{i \Delta x \, \sigma_i \omega_i^E/2} - e^{-i \Delta x \, \sigma_i
\omega_i^E/2}) \choose \sinh \pi P_i
(e^{i \Delta x \, \sigma_i \omega_i^E/2} + e^{-i \Delta x\, \sigma_i \omega_i^E/2}) }   \\ &&
\left[\exp\left\{ {2\over \pi} \int {d^2 k_E\over k_E^2} \left(1-e^{-\pi k_E^2 /\eps} \right)
\left\{ e^{-i\omega^E \Delta x} (\cosh^2 (\pi P) + \sinh^2(\pi P)) 
\right\} \right\}-1\right]\, . \nonumber
\een
To go from the first expression to the second expression, we change variables to
$x=(x_1+x_2)/2$, $\Delta x=(x_1-x_2)$ and then do the $x$ integral by 
changing variables to $x=i y$, $\omega_i^E=-i\omega_i$. This gives a factor of
$2\pi i \delta\left(\sum_i \sigma_i\omega_i\right)$.
Due to the $(1-e^{-\pi k_E^2 /\eps})$ factor in the integrand that vanishes
at $k_E^2=0$, this
vertex has no singularity from $k_E^2=0$
even when the external momenta are Lorentzian, i.e. when the
$\omega_i$'s are real. This justifies declaring this as a single composite vertex.
The apparent ultraviolet divergence of the $k_E$ integral in the last line can be
avoided for $|\Delta x|>2\pi$ by doing the $\omega^E$ integration before the $P$
integration. The result will have a singularity at $\Delta x=\pm 2\pi$ and the integrand has
be continued to $|\Delta x|<2\pi$ via analytic continuation. If we want the interaction
vertex \refb{e3.17}
to correspond to a real term in the effective action, then we need to use the
`unitary prescription' for integrating over $\Delta x$\cite{2012.00041},
which in this case translates to the
principal value prescription.
\item We shall define the propagator of a closed string of momentum $k=(\omega,P)$
to be
\be
-{8\pi i\over k^2-i\ve} = {8\pi i\over \omega^2 - P^2 +i\ve}\, ,
\ee
and take the integration measure over the internal momenta to be 
$d^2 k/(4\pi^2)$, so that
\be
{d^2 k\over 4\pi^2} \, \left(-{8\pi i\over k^2-i\ve}\right)
= {2\over \pi} {d^2 k_E 
\over k_E^2}\, .
\ee
\end{enumerate}

\begin{figure}
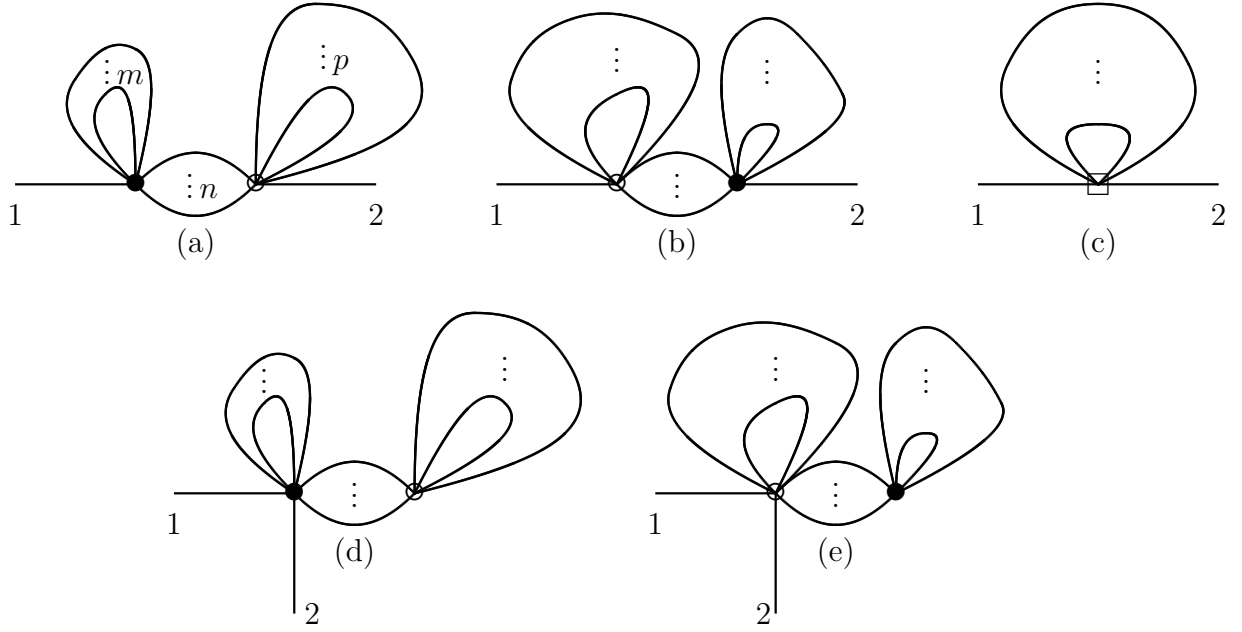


\begin{center}

\figone

\vskip -.6in

\hbox{~\hskip .9in \figtwo}

\end{center}

\vskip -1in

\caption{Feynman diagram representation of \refb{e3.14}.
\label{f1}}

\end{figure}

We now claim that $\MM_2(\omega_1,\omega_2)$ given in \refb{e3.14}
can be regarded as a sum of 
contributions from the Feynman diagrams shown in Fig.~\ref{f1}.  
Since the external incoming and outgoing states are NS+R sector scalars, for the
D-instanton vertex given in \refb{e3.15} they
couple via terms proportional to $\cosh(\pi P_1)+\sinh(\pi P_1)=e^{\pi P_1}$ for
the incoming state and $-\cosh(\pi P_2)+\sinh(\pi P_2)=-e^{-\pi P_2}$ for the outgoing
state. On the other hand, for the anti-D-instanton vertex given in \refb{e3.16}, they
couple via terms proportional to $-\cosh(\pi P_1)+\sinh(\pi P_1)=-e^{-\pi P_1}$ for
the incoming state and $\cosh(\pi P_2)+\sinh(\pi P_2)=e^{\pi P_2}$ for the outgoing
state.
Therefore the contribution
from Fig.~\ref{f1}(a) is given by:
\ben\label{e3.19}
&& e^{-2 S_D} {\eps\over 16\pi^3} e^{\gamma_E} {1\over n!m!p!}
\int \prod_{i=1}^n {d\tilde\omega_i\over 2\pi} {d\tilde P_i\over 2\pi} {8\pi i\over -\tilde k_i^2 +i\ve}
e^{-\pi \tilde k_i^2/\eps} \, 2\pi i \, \delta\Big(\omega_1 - \sum_{i=1}^n \tilde \omega_i\Big) 
2\pi i\, \delta\Big(\omega_2
- \sum_{i=1}^n \tilde \omega_i\Big) \nonumber \\ &&
\hskip 2in \left\{\cosh^2(\pi \tilde P_i) + \sinh^2(\pi \tilde P_i)\right\}\,
e^{\pi P_1+\pi P_2} 
 \\
&& \left(\int \prod_{i=1}^m \left( - {1\over 2}\right)
{d\hat\omega_i\over 2\pi} {d\hat P_i\over 2\pi} {8\pi i\over -\hat k_i^2 +i\ve}
e^{-\pi \hat k_i^2/\eps}\right) \, \left( 
\int \prod_{i=1}^p \left( - {1\over 2}\right)
{d\bar\omega_i\over 2\pi} {d\bar P_i\over 2\pi} {8\pi i\over -\bar k_i^2 +i\ve}
e^{-\pi \bar k_i^2/\eps}\right)\, . \nonumber
\een
Note the factors of $1/2$ in the integrands in the last line -- these are the
correct combinatoric factors associated with the propagators with both ends on the same
vertex. The minus sign comes from the vertex factors, since an internal line
of momentum $(\omega,P)$ joining the same vertex  generates $-\cosh^2(\pi P)
+ \sinh^2(\pi P)=-1$.

\refb{e3.19} can be identified to the following contribution from \refb{e3.14}:
\begin{enumerate}
\item Pick the term
\be\label{enew3.14}
e^{\pi P_1+\pi P_2} e^{i \omega_1^E x_1 - i\omega_2^E x_2} 
\ee
from the last line of \refb{e3.14}.
\item Pick $n$ factors of 
\be
{2\over \pi} \int {d^2 k_E\over  k_E^2} e^{-\pi k_E^2 /\eps} 
\left\{ e^{-i\omega^E (x_1-x_2)} (\cosh^2 (\pi P) + \sinh^2(\pi P)) 
\right\}
\ee
from the expansion of the exponential in the second line of \refb{e3.14}.
\item Pick $m$ factors of 
\be
-{1\over \pi} \int {d^2 k_E\over k_E^2} e^{-\pi k_E^2 /\eps} 
\ee
from the expansion of the first term
in the third line of \refb{e3.14}.
\item Pick $p$ factors of 
\be
-{1\over \pi} \int {d^2 k_E\over  k_E^2} e^{-\pi k_E^2 /\eps} 
\ee
from the expansion of the second term
in the third line of \refb{e3.14}.
\item Pick 1 in the expansion of the exponential in the fourth line of \refb{e3.14}.
\end{enumerate}
The integrations over $x_1$ and $x_2$ in \refb{e3.14}
generate the two energy conserving delta functions
in \refb{e3.19}.  For this we need to rotate the $\omega_i$'s from the imaginary axis
to the real axis via the first quadrant and rotate the $x_1,x_2$ integration contours in the
opposite direction so as to keep $\omega x$ fixed. This
produces the factors of $i$ multiplying the
delta functions. Also $\int d^2 k_E/k_E^2$ becomes $d^2 k (-i/k^2)$ in Lorentzian variables.

The contributions from Fig.~\ref{f1}(b),(d) and (e)
can be interpreted in the same way, except that from the  last line of \refb{e3.14} we pick
respectively the terms proportional to $e^{-\pi(P_1+P_2)}$, $e^{\pi(P_1-P_2)}$ and
$e^{\pi(P_2-P_1)}$  instead of \refb{enew3.14}. 
Therefore the sum of these diagrams produces all
the terms in the expansion of \refb{e3.14}, other than those obtained from the higher
order terms in the expansion
of the exponential in the penultimate line of \refb{e3.14}. The contribution from
Fig.~\ref{f1}(c) produces this contribution.\footnote{Note that the contribution from
individual Feynman diagrams diverge in the $\eta\to 0$
limit. However, the finiteness of the original expression implies that the sum of all the
Feynman diagrams with a fixed number of total propagators is infrared finite. We shall
implicitly follow this procedure even when some of the propagators are cut, summing 
over all graphs with a fixed number of total propagators.
The infrared finite expression
\refb{e2.7rep}
that we get at the end for sum over cut diagrams should be regarded as a result of
organizing the sum over diagrams this way.}

\begin{figure}
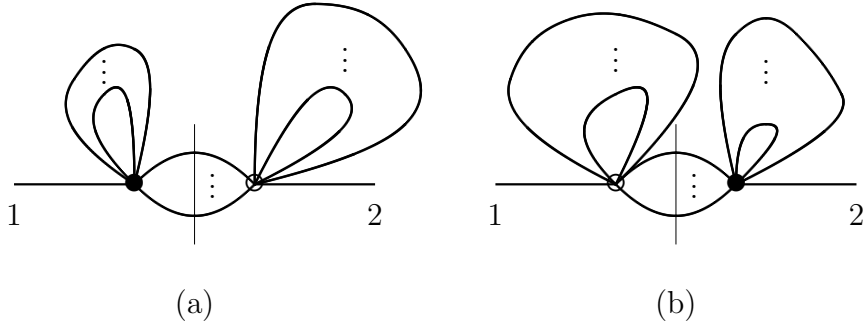

\begin{center}

\hbox{~\hskip 1in \figthree}

\end{center}

\vskip-1.2in

\caption{Contributions to $\MM_1^* \MM_1$. The thin vertical lines represent cuts.
\label{f2}}

\end{figure}

The Cutkosky rules tell us that the sum over all the cuts of the Feynman diagrams of
Fig.~\ref{f1} vanishes. Of these the cuts that are fully to the left or fully to the right of the
diagrams, without cutting any internal propagators, produce the factor of $\MM_2
+\MM_2^*$.
The other cuts are those of Fig.~\ref{f1}(a) and (b), cutting all the internal propagators
that connect the D-instanton induced vertex to the anti-D-instanton induced vertex. 
These have been shown in Fig.~\ref{f2}(a) and (b) respectively.
This
gives $\MM_1^* \MM_1$, producing the unitarity relation $\MM_1^* \MM_1 =-
(\MM_2+\MM_2^*)$. Therefore as long as Cutkosky rules are applicable, \refb{e2.7}
should hold. 
The original proof of Cutkosky rules using largest time equation\cite{veltman,diagrammar}
and a different
version given in perturbation theory\cite{sterman} do not hold for non-local vertices,
containing exponential in momenta, that
we have in string field theory  e.g. the 
$e^{-\pi k^2/2\eps}$ factors in \refb{e3.15}-\refb{e3.17}. A perturbative proof that holds for these
cases  was given in \cite{1604.01783}. Since the D-instanton induced vertices are of the same
type, we can still make use of the proof given in \cite{1604.01783}.
Furthermore, the proof of Cutkosky rules
given in \cite{1604.01783} used manipulations of energy integration contour at fixed values
of spatial components of the loop momenta. Therefore putting a lower cut-off on the Liouville
momentum does not affect the proof of Cutkosky rules.
The final ingredient
in the proof was the reality of the action. The sum of D-instanton and anti-D-instanton
induced vertices is manifestly real, but the reality of the composite vertex  \refb{e3.17}
requires that we use the unitary prescription for integrating over
$\Delta x = x_1-x_2$\cite{2012.00041}.
Once this is done, the Cutkosky rules hold. Notwithstanding these general arguments,
we shall now
explicitly verify that the 
expression for $\MM_1^* \MM_1$ computed from the sum over cuts reproduces
\refb{e2.7}.  Besides providing a check on the abstract arguments of
\cite{1604.01783,2012.00041} for the
validity of the cutting rules, this will be useful in computing the semi-inclusive cross 
section where in the sum over
states on the left hand side of \refb{e2.7} we sum
over only a subset of states.

The cut diagram shown in Fig.~\ref{f2}(a), obtained by replacing the $i/(-\tilde k_i^2+i\ve)$
by $2\pi \delta(\tilde k_i^2)\Theta(\tilde\omega_i)$, and complex conjugating the contribution
from the right of the cut, is given by:
\ben\label{e3.24}
&&\hskip -.2in
 e^{-2 S_D} {\eps\over 16\pi^3} e^{\gamma_E} \sum_{n=1}^\infty \sum_{m,p=0}^\infty
{1\over n!m!p!}
\int \prod_{i=1}^n \left\{4{d\tilde\omega_i} {d\tilde P_i}  \delta(\tilde k_i^2)
\Theta(\tilde\omega_i)
\left(\cosh^2(\pi \tilde P_i) + \sinh^2(\pi \tilde P_i)\right)\right\} \nonumber \\ &&
\hskip 2in \, (2\pi i) \delta\left(\omega_1 - \sum_{i=1}^n \tilde \omega_i\right) (-2\pi i )\delta\left(\omega_2
- \sum_{i=1}^n \tilde \omega_i\right) 
 \\
&&\hskip -.2in e^{\pi P_1+\pi P_2}  \int \prod_{i=1}^m  \left\{ \left( - {1\over \pi}\right)
{d\hat\omega_i} {d\hat P_i} {i\over -\hat k_i^2 +i\ve}
e^{-\pi \hat k_i^2/\eps}\right\} \, 
\int \prod_{i=1}^p \left\{  \left( - {1\over \pi}\right)
{d\bar\omega_i} {d\bar P_i} {i\over -\bar k_i^2 +i\ve}
e^{-\pi \bar k_i^2/\eps}\right\}^*\, . \nonumber
\een
The sum over $m$ and $p$ easily exponentiates to
\be\label{e3.24aa}
\exp\left[- {2\over \pi} \int {d^2 k_E\over  k_E^2} e^{-\pi k_E^2 /\eps} \right]\, .
\ee
Defining $(u,v) = (P\sqrt{\pi/\eps}, \omega^E \sqrt{\pi/\eps})$ and noting that the limits
of integration are $\eta\le P<\infty$, $-\infty<\omega^E<\infty$, we can express \refb{e3.24aa}
as\footnote{The constant term in the exponent of \refb{e3.26}
was found numerically. At present we do not
have an analytic derivation of this term.}
\ben\label{e3.26}
\exp\left[- {2\over \pi} \int_{\eta\sqrt{\pi/\eps}}^\infty du \int_{-\infty}^\infty dv 
{e^{-(u^2+v^2)}\over u^2 + v^2}  \right] &=& \exp\left[ \ln {\pi\eta^2\over \eps} +\gamma_E+2\ln 2 +\OO(\eta/\sqrt{\eps})
\right] \nonumber \\
&=& {4\pi\eta^2\over \eps} e^{\gamma_E} \left(1 + \OO(\eta/\sqrt{\eps})\right) ,
\een
where we have organized the expansion so that we take the $\eta\to 0$ limit before taking the
$\eps\to 0$ limit.

The sum over $n$ in \refb{e3.24} takes the form:
\ben
&& 2\pi \delta(\omega_1-\omega_2) f(\omega_1), \nonumber \\
f(\omega) &\equiv& \sum_{n=1}^\infty {1\over n!} \int \prod_{i=1}^n \left\{
{2d\tilde P_i\over \tilde P_i}\, (\cosh^2(\pi \tilde P_i) +\sinh^2(\pi \tilde P_i)) \right\} \, 
2\pi \delta\Big(\sum_{i=1}^n \tilde P_i - \omega\Big)\, .
\een
To find $f(\omega)$ we first compute\cite{2012.00041,2204.01747}
\ben 
\int_0^\infty e^{-\nu\omega} f(\omega) {d\omega \over 2\pi} &=&
 \sum_{n=1}^\infty {1\over n!} \int \prod_{i=1}^n \left\{
{2d\tilde P_i\over  \tilde P_i}\, (\cosh^2(\pi \tilde P_i) +\sinh^2(\pi \tilde P_i))
e^{-\nu \tilde P_i}\right\} \nonumber \\
&=& \exp\left[\int {2d\tilde P\over \tilde P}\, 
(\cosh^2(\pi \tilde P) +\sinh^2(\pi \tilde P))  e^{-\nu \tilde P} \right]-1\, .
\een
Recalling that the lower limit of $\tilde P$ integration is $\eta$, we get,
\be
\int_0^\infty e^{-\nu\omega} f(\omega) {d\omega \over 2\pi} =
\exp[-2\gamma_E - \ln \{\eta^2 (\nu^2-4\pi^2)\}+\OO(\eta)]-1
=e^{-2\gamma_E} {1\over \eta^2 (\nu^2-4\pi^2)} + \OO(1/\eta) \, .
\ee
Demanding that this holds for all $\nu$, we get
\be\label{e3.30}
f(\omega) = {e^{-2\gamma_E}\over \eta^2} \sinh(2\pi \omega) + \OO(1/\eta)\, .
\ee
Substituting \refb{e3.26} and \refb{e3.30} into \refb{e3.24}
we get the following expression for Fig.\ref{f2}(a):
\be\label{enew3.25}
2\pi \delta(\omega_1-\omega_2)\, 
e^{-2 S_D} {\eps\over 16\pi^3} e^{\gamma_E} \,  {4\pi\eta^2\over \eps} \, 
e^{\gamma_E}\, {e^{-2\gamma_E}\over \eta^2} \sinh(2\pi \omega_1) \, e^{2\pi\omega_1}\, ,
\ee
where we have used $P_1=\omega_1=\omega_2=P_2$. The contribution from 
Fig.\ref{f2}(b) has a similar form except that we
have $e^{-2\pi\omega_1}$
instead of $e^{2\pi\omega_1}$. Taking the sum of the two cut diagrams we get:
\be\label{e2.7rep}
{\sum_n} \MM_1(\omega_1,n) \MM_1(\omega_2,n)^* =
\delta(\omega_1-\omega_2)\, 
e^{-2 S_D} \, {1\over \pi}  \,  \sinh(2\pi \omega_1) 
\, \cosh(2\pi\omega_1)\, .
\ee
This agrees with \refb{e2.7}.

\sectiono{Semi-inclusive cross section} \label{s4}

Our goal in this section will be to find the D-instanton induced semi-inclusive
cross section for a right sector closed string state of energy $\omega_1$ to go into a
set of $r$  right sector closed
string states of energies in the range $(e_1, e_1+\Delta e_1),\cdots, 
(e_r, e_r+\Delta e_r)$,
$l$ left sector closed
string states of energies in the range $(e'_1, e'_1+\Delta e'_1),\cdots, 
(e'_l, e_l'+\Delta e'_l)$
and any number of other closed string states. $\Delta e_i$ and $\Delta e_i'$ are taken to
be infinitesimal.
To calculate this, we reexamine the expression \refb{e3.24} of the cut diagram
associated with Fig.~\ref{f2}(a). Recalling the couplings of NS and R sector states to
the D-instanton and anti-D-instanton induced vertices given in \refb{e3.15} and \refb{e3.16}
respectively, and recalling that the right and left sector closed string states are
given respectively by the sum and difference of the NS and R sector states,
one finds that a right sector state propagating from the left to the right of the cut in
Fig.\ref{f2}(a) will have coupling proportional to $e^{-2\pi \tilde P_i}$ and
a left sector state propagating from the left to the right of the cut in
Fig.\ref{f2}(a) will have coupling proportional to $e^{2\pi \tilde P_i}$.
Writing $\cosh^2 (\pi \tilde P_i) + \sinh^2(\pi \tilde P_i)$ as $(e^{2\pi \tilde P_i}+e^{-2\pi \tilde P_i})/2$ in
\refb{e3.24} we see that the $e^{2\pi \tilde P_i}/2$ factor can be traced to the propagation of
a left-sector closed string and the $e^{-2\pi \tilde P_i}/2$ factor can be traced to the propagation
of a right sector closed string. Therefore to compute the desired semi-inclusive cross section,
we need to,
\begin{enumerate}
\item
restrict the integration over $r$ of the $\tilde\omega_i$'s
in \refb{e3.24} to the range $(e_1, e_1+\Delta e_1),\cdots, 
(e_r, e_r+\Delta e_r)$ and replace the  $\cosh^2 (\pi \tilde P_i) + \sinh^2(\pi \tilde P_i)$
factor for these momenta by $e^{-2\pi \tilde P_i}/2=e^{-2\pi e_i/2}$, 
\item restrict the integration over $l$ of the other $\tilde\omega_i$'s
in \refb{e3.24} to the range $(e'_1, e'_1+\Delta e'_1),\cdots, 
(e'_l, e'_l+\Delta e'_l)$ and replace the  $\cosh^2 (\pi \tilde P_i) + \sinh^2(\pi \tilde P_i)$
factor for these momenta by $e^{2\pi \tilde P_i}/2=e^{2\pi e'_i}/2$, 
\item and let the rest of the integrals run over the full range.
\end{enumerate}
Also since there are $n! / (n-r-l)!$ ways of choosing the $r+l$ variables among the $n$
integration variables $\tilde \omega_i$ whose integration ranges are restricted to 
$(e_1, e_1+\Delta e_1),\cdots, 
(e_r, e_r+\Delta e_r)$, $(e'_1, e'_1+\Delta e'_1),\cdots, 
(e'_l, e'_l+\Delta e'_l)$, we get an extra multiplicative factor of 
$n! /  (n-r-l)!$ that converts
the $1/n!$ in \refb{e3.24} to $1/ (n-r-l)!$. 
We can now perform the sum over
$n$, $m$ and $p$ as before, and the result takes the form of \refb{enew3.25} with
$\omega_1$ replaced by $\omega_1 - \sum_i e_i - \sum_i e'_i$ in the argument of
$\sinh(2\pi\omega_1)$, 
multiplied by a factor of $\Delta e_i e^{-2\pi e_i}/e_i$ for each
final state right sector closed string and a factor of $\Delta e'_i e^{2\pi e'_i}/e'_i$ for
each final state left sector closed string.\footnote{Since the integration measure in the
sum over states is $de/e$, this justifies the definition of the identity matrix as $\omega_1
\delta (\omega_1-\omega_2)$ as in \refb{e2.5aa}.} This gives
\ben
&& 2\pi \delta(\omega_1-\omega_2)\, 
e^{-2 S_D} {1\over 4\pi^2}  \,  \sinh\left(2\pi \left(\omega_1
- \sum_{i=1}^r e_i -\sum_{i=1}^l e'_i\right)\right) \, e^{2\pi \omega_1}\nonumber \\
&& \hskip 1in \times
 \left\{\prod_{i=1}^r {\Delta e_i\, e^{-2 \pi e_i} \over e_i} \right\}\, 
\left\{\prod_{i=1}^l {\Delta e'_i e^{2\pi e_i'}\over e'_i} \right\}\, .
\een
Similarly the contribution to this semi-inclusive cross section from the cut diagram of
Fig.\ref{f2}(b) is given by:
\ben
&& 2\pi \delta(\omega_1-\omega_2)\, 
e^{-2 S_D} {1\over 4\pi^2}  \,  \sinh\left(2\pi \left(\omega_1
- \sum_{i=1}^r e_i -\sum_{i=1}^l e'_i\right)\right) \, e^{-2\pi \omega_1}\nonumber \\
&& \hskip 1in \times
\left\{\prod_{i=1}^r {\Delta e_i\, e^{2 \pi e_i} \over e_i} \right\}\, \left\{\prod_{i=1}^l {\Delta e'_i e^{-2\pi e_i'}\over e'_i} \right\}\, .
\een
Adding these two contributions we get,
\ben \label{efinexp}
&& {\sum_n}' \MM_1(\omega_1,n) \MM_1(\omega_2,n)^* \nonumber \\
&=& e^{-2S_D}\, \left\{\prod_{i=1}^r {\Delta e_i\over e_i} \right\}\, \left\{\prod_{i=1}^l {\Delta e'_i\over e'_i} \right\}\, 
 \delta\left(\omega_1 -\omega_2 \right)\, {1\over \pi} \,
\sinh\left(2\pi \left(\omega_1
- \sum_{i=1}^r e_i -\sum_{i=1}^l e'_i\right)\right) \nonumber \\ && \hskip 1in \times
\cosh\left(2\pi \left(\omega_1+ \sum_{i=1}^l e'_i - \sum_{i=1}^r e_i\right)\right)\, ,
\een
where $\sum'$ on the left hand side denotes sum over all final states that contain 
$r$  right sector closed
string states of energy in the range $(e_1, e_1+\Delta e_1),\cdots, 
(e_r, e_r+\Delta e_r)$,
$l$ left sector closed
string states of energy in the range $(e'_1, e'_1+\Delta e'_1),\cdots, 
(e'_l, e_l+\Delta e'_l)$
and any number of other closed string states, with the restriction
$\omega_1 > \sum_i e_i + \sum_i e'_i$.

\sectiono{Matrix model computation} \label{s5}

We shall now see how to compute the semi-inclusive cross section in the matrix model.
The computation in this case simplifies by noting that in the semi-inclusive cross section, the
sum over `anything else' can be taken in the fermionic basis, since the free fermions
and holes form a complete basis of states. 

We shall first illustrate this procedure by computing the
contribution to the fully inclusive cross section
induced by single instanton or single anti-instanton\cite{0309148}.
Since single instanton induces transmission of a fermion or
a hole, and since the incoming closed string is a fermion hole pair, we can compute
the inclusive cross section by summing over two final states: (1) the fermion is
transmitted and the hole is reflected back and (2) the hole is transmitted and the
fermion is reflected back. Let us denote by $e'$ and $e$ the energies of
the transmitted and the
reflected particle respectively in string units. Using the convention of \cite{2204.01747}
that the energy interval $e$ in string theory corresponds to energy interval
$2e$ in the matrix model,
and that $-\mu$ is the fermi energy of the matrix model, we see that a fermion carrying
energy $e$ in string units 
has energy $-\mu + 2e$ in the matrix model and a hole carrying energy $e$ in string
units correspond to a hole at energy level $-\mu-2e$ in the matrix model.
If we denote the $T(x)$ and $R(x)$ the reflection and transmission coefficient of
a fermion carrying energy $x$ in the matrix model, then the net contribution to the
$\sum_{n}\MM_1(\omega_1,n) \MM_1(\omega_2,n)^*$ is given by:\footnote{There is no
energy dependent normalization in the phase space integration measure for non-relativistic
fermions. To check the overall normalization in \refb{etotal}, we note that the leading identity
matrix in $S^\dagger S$ comes from the term where both the fermion and the hole are reflected
and we approximate the reflection coefficient $R$ by 1. In this case the second line of
\refb{etotal} would be replaced by 1 and the  integral in the first line gives $\omega_1
 \delta(\omega_1-\omega_2)$
in agreement with \refb{e2.5aa}.}
\ben\label{etotal}
&& \int_0^\infty {de\over 2\pi} \int_0^\infty {de'\over 2\pi} \, 2\pi\delta(e+e'-\omega_1) \, 
2\pi \delta(e+e'-\omega_2) \nonumber\\
&& \hskip 1in  \times \,
\left[ |T(-\mu+2e') R(-\mu-2e)^*|^2 + |T(-\mu-2e')^* R(-\mu+2e)|^2\right]\, .
\een
The first term in the square bracket represents the contribution where the transmitted
particle is a fermion, while the second term represents the contribution where the
transmitted particle is a hole. In writing \refb{etotal} we have used the result that
the reflection and transmission coefficients of a hole are given by the complex conjugates
of those of the fermion. To this order we can approximate $T(x)$ and $R(x)$ up to a
phase by
\be \label{etransrefl}
T(x) \simeq e^{\pi x}, \qquad R(x) \simeq 1\, .
\ee
Substituting this into \refb{etotal} we get,
\be
\delta(\omega_1-\omega_2)\, {1\over \pi}\, e^{-2\pi\mu} \, \sinh(2\pi\omega_1)\cosh(2\pi\omega_1)\, .
\ee
This agrees with the string theory result \refb{e2.7rep} once we identify $e^{-2S_D}$ with
$e^{-2\pi\mu}$. From now on `energy' will always be understood as the energy measured
in string units unless mentioned otherwise.

\begin{figure}
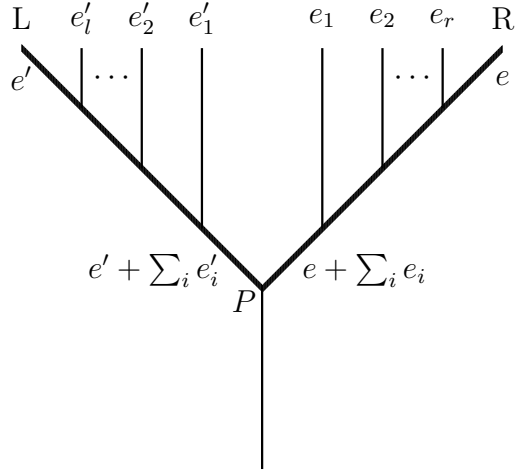


\begin{center}

\figfive

\end{center}

\vskip -.3in

\caption{Diagrammatic representation of a scattering process in which an incoming
closed string state carrying energy $\omega_1$ splits into a set of right sector closed
strings carrying energies $e_1,\cdots , e_r$, a set of left sector closed
strings carrying energies $e'_1,\cdots , e'_l$ and a fermion hole pair, one on each side
of the potential barrier. The time flows up, the thin lines denote the
external closed strings and the thick lines denote fermions and holes. There are
two distinct diagrams, one where the left sector state is a fermion and the right
sector state is a hole and vice versa.\label{f5}}

\end{figure}

Next we consider the case of semi-inclusive cross section where the final state contains
$r$  right sector closed
string states of energy in the range $(e_1, e_1+\Delta e_1),\cdots, 
(e_r, e_r+\Delta e_r)$,
$l$ left sector closed
string states of energy in the range $(e'_1, e'_1+\Delta e'_1),\cdots, 
(e'_l, e'_l+\Delta e'_l)$ plus any other state. We can choose the basis of
`any other states' as fermion or hole states. 
We can compute this amplitude with the help of real time diagrams introduced in
\cite{9111035}, except that here in the final state we allow free fermions and holes represented by
open lines. 
A diagrammatic representation of
the process under consideration
is shown in Fig.~\ref{f5}.  The basic process is that the initial closed
string, containing the fermion hole pair, splits into a fermion and a hole, one on either side
of the potential barrier, which then rearrange themselves to a set of closed strings
and the final state fermion hole pair on the opposite sides of the barrier. 
The only non-trivial
part of this diagram is the `interaction vertex' $P$, 
given by the product of the transmission coefficient of a fermion of energy
$e'+\sum_i e'_i$ and reflection coefficient of a hole of energy $e+\sum_i e_i$
or vice versa. Leaving out the phase space factors $\Delta e_i/e_i$ and $\Delta e_i'/e_i'$,
which have the same origin in string theory and the matrix model, we see from Fig.~\ref{f5}
that the contribution is given by an expression similar to \refb{etotal} with $e$ replaced by
$e+\sum_i e_i$ and $e'$ replaced by $e'+\sum_i e'_i$ in the integrand:
\ben
&& \int_0^\infty {de\over 2\pi} \int_0^\infty {de'\over 2\pi} \, 2\pi\delta\left(e+e'-\omega_1
+\sum_i e_i + \sum_i e'_i\right) \, 
2\pi \delta\left(e+e'-\omega_2+\sum_i e_i + \sum_i e'_i\right) \nonumber\\
&&  \times \,
\left[ \left|T\left (-\mu+2e' +2\sum_i e'_i\right) R\left(-\mu-2e-2\sum_i e_i\right)^*\right|^2 \right.
\nonumber \\ && \left. \qquad + \, \left|T\left(-\mu
-2e' - 2\sum_i e'_i\right)^* R\left(-\mu+2e+2\sum_i e_i\right)\right|^2\right]\, .
\een
Using \refb{etransrefl} we can reduce this to,
\be
 \delta(\omega_1-\omega_2) \, 
{1\over \pi} \, e^{-2\pi\mu} \, \sinh\left(2\pi\left(\omega_1 -\sum_i e_i - \sum_i e'_i\right)\right)\cosh\left(2\pi\left(\omega_1-\sum_i e_i + \sum_i e'_i\right)\right)\, .
\ee
This is in perfect agreement with \refb{efinexp}.

\sectiono{Discussion} \label{s6}

Infrared divergences in the two dimensional type 0B string theory and those
in four dimensional quantum electrodynamics share many common features. In both
cases the infrared divergences make the usual S-matrix vanish, but the semi-inclusive
cross section, where we allow in the final state arbitrary number of soft particles, is
finite. In type 0B string theory the infrared divergence can be traced to the fact that the
final state may be a state containing fermion hole pair on opposite sides of the potential
barrier, and this cannot be described as a collection of finite number of closed strings.
Put another way, the final state may be in a different  `charge sector'
compared to the initial state\cite{0309148,2204.01747}. In quantum electrodynamics
the vanishing of the S-matrix due to infrared divergence can be traced to the fact that
the final state after the scattering is built on a different vacuum
compared to the initial state\cite{1705.04311}.  In type 0B string theory the infrared
divergence in the S-matrix can be cured by allowing the final state to have a fermion
hole pair on opposite sides of the potential barrier. In quantum electrodynamics
the infrared divergences can be cured by using the  Faddeev - Kulish
states\cite{chung,kibble,Kulish:1970ut}.
This suggests that in quantum
electrodynamics, the analog of the
state containing fermion hole pair on opposite sides of the potential barrier may be related
to the photon cloud in the Faddeev - Kulish states, together with a finite number of
photons to balance energy and momentum.  
It will be interesting to  explore this analogy in more detail.

\bigskip 

\noindent{\bf Acknowledgement}: We wish to thank  Bruno Balthazar, 
Joydeep Chakravarty, Alok Laddha, Victor Rodriguez and
Xi Yin for useful discussions at various stages of this work. 
This work is supported by ICTS-Infosys Madhava 
Chair Professorship
and the J.~C.~Bose fellowship of the Department of Science and Technology,

\end{document}